# Technology Integration in the Project Based Learning Model: Bibliometric Analysis 2015-2024

Intan Purnama Yani [1] , Serli Ahzari [1] , Asrizal [1*], Fuja Novitra[1]

[1] *Department of Physics, Universitas Negeri Padang, Jl. Prof. Dr. Hamka, Air Tawar Barat, Padang 25131, Indonesia*
*Corresponding author. asrizal@fmipa.unp.ac.id*

**ABSTRACT**

*This study aims to conduct a bibliometric analysis of scientific publications discussing technology integration in the Project-Based Learning (PjBL) model during the 2015-2024 period. With the evolving 21st-century educational paradigm, PjBL has emerged as one of the most promising pedagogical approaches. In this context, the integration of technology in PjBL becomes an important catalyst that expands the scope and depth of learning experiences while preparing students to face challenges in the digital era. This research applies a bibliometric analysis methodology and systematic literature review, utilizing the Publish or Perish (POP) application for data collection and VOSviewer for data analysis and visualization. The results reveal emerging research trends, collaboration patterns, and focus areas in the scientific literature related to technology integration in PjBL. The bibliometric analysis uncovers research dynamics and knowledge gaps that need to be addressed. These findings provide valuable insights for researchers, practitioners, and policymakers in optimizing the implementation of technology-based PjBL and formulating educational policies responsive to technological developments and 21st-century learning needs. The results of this study are expected to contribute significantly to the academic community and educational practitioners in developing more effective technology-based PjBL implementation strategies and fostering interdisciplinary collaboration to advance education in the digital age.*

## I. INTRODUCTION

The 21st century educational paradigm has undergone a significant transformation, with Project Based Learning (PjBL) emerging as one of the most promising pedagogical approaches. PjBL offers a framework that allows students to engage in active, collaborative, and real problem-oriented learning [1]. In this context, the integration of technology in PjBL becomes an important catalyst that expands the scope and depth of learning experiences, while preparing students to face challenges in the digital era [2]. The evolution of information and communication technology (ICT) has opened up new opportunities in PjBL implementation. Technology not only functions as a supporting tool, but also as an integral component that changes the nature of the learning project itself [3]. The integration of technology in PjBL allows students to access global resources, collaborate without geographic boundaries, and develop digital literacy that is crucial for future success [4].

In Indonesia, the adoption of technology-based PjBL has gained momentum, in line with the government's vision to improve the quality of national education through the use of ICT [5]. However, effective implementation of this approach still faces various challenges, including the digital divide, teacher preparedness, and uneven technology infrastructure. A comprehensive understanding of the research landscape regarding technology integration in PjBL is critical. Bibliometric analysis offers a systematic methodology for mapping and evaluating research developments in this field [6]. Through this approach, we can identify trends, collaboration patterns, and focus areas in the scientific literature, thereby providing valuable insights for researchers, practitioners, and policy makers [7]. In the context of technology integration in PjBL, bibliometric analysis can reveal research dynamics and knowledge gaps that need to be addressed.

Previous studies have shown that the integration of technology in PjBL can increase learning motivation, problem solving skills, and student learning outcomes [8]. Technologies such as virtual simulations, augmented reality, and online collaboration platforms have been shown to enrich the PjBL experience by enabling the exploration of complex concepts and deeper interactions [9]. However, the effectiveness of this integration depends on various factors, including appropriate learning design, adequate technological support, and the readiness of all stakeholders [10]. In a global context, research on technology integration in PjBL has shown rapid development. International comparative studies reveal variations in approaches and outcomes, which are influenced by cultural factors, educational policies, and technological infrastructure [11]. On the other hand, research in Indonesia shows great potential as well as unique challenges in implementing technology-based PjBL, especially related to geographic and socio-economic diversity.

Despite this, comprehensive syntheses of research in this field, especially those that include both global and local perspectives, are still limited. Bibliometric analyzes covering the period 2015–2024 can provide a clearer picture of the evolution of research, identify knowledge gaps, and guide future research agendas [12]. This understanding is fundamental to optimize technology integration in PjBL and improve the quality of education in the digital era. This research aims to conduct a bibliometric analysis of scientific publications discussing technology integration in the Project Based Learning model during the 2015-2024 period. The results of this study are expected to provide significant contributions to the academic community and educational practitioners. By identifying emerging and underexplored research areas, this study can inform future research directions and encourage interdisciplinary collaboration. For practitioners and policy makers, these findings can be the basis for developing more effective technology-based PjBL implementation strategies, as well as formulating educational policies that are responsive to technological developments and 21st century learning needs [13]. This is very important considering the strategic role of education in preparing Indonesia's young generation to compete in a global era that is increasingly technology-based [14].

The integration of technology in PjBL has become increasingly relevant in the context of the COVID-19 pandemic, which has accelerated the adoption of digital learning platforms and tools [15]. As educational institutions worldwide were forced to shift to remote learning, the need for effective pedagogical approaches that leverage technology became more pressing than ever [16]. PjBL, with its emphasis on student-centered, inquiry-based learning, has shown promise in adapting to these new circumstances [17]. However, the rapid transition to online learning has also highlighted the digital divide and the challenges of ensuring equitable access to technology-enhanced education [18]. A comprehensive analysis of research on technology integration in PjBL during this period can shed light on how educators and researchers have responded to these challenges and provide valuable insights for post-pandemic education.

Moreover, the growing emphasis on 21st-century skills, such as critical thinking, collaboration, and digital literacy, has further underscored the importance of technology integration in PjBL [1]. As the world becomes increasingly interconnected and technology-driven, it is crucial for students to develop the competencies needed to thrive in a rapidly changing global landscape [19]. PjBL, with its focus on real-world problem-solving and authentic learning experiences, provides an ideal framework for fostering these skills [20]. By examining the research on technology integration in PjBL, this study can contribute to a deeper understanding of how technology can be leveraged to support the development of 21st-century skills and prepare students for the challenges and opportunities of the future.

## II. METHOD

This research applies a bibliometric analysis methodology and systematic literature review, focusing on the theme of technology integration in Project Based Learning (PjBL). In its implementation, this research utilized two main tools: Publish or Perish (POP) for data collection, and VOSviewer for data analysis and visualization. VOSviewer was chosen for its ability to produce visual representations of bibliometric maps and identify publication trends over a period of time. The research process consists of several key stages: first, data extraction via the POP application; second, systematic documentation of search results from POP; third, processing and analyzing article data using VOSviewer; fourth, storing analysis results in visual format; and finally, an in-depth interpretation of the relationships between the identified concepts.

The analysis results produced by VOSviewer can be categorized into three main types of visualization. Network visualization illustrates the interconnections between key concepts in research on technology integration in PjBL. The overlay visualization charts the evolution of research over the past decade, showing shifts in focus and trends. Meanwhile, the density visualization indicates the intensity of use and relevance of certain terms in the context of this research, highlighting areas that have received significant attention in the literature. This research aims to provide a comprehensive understanding of the current research landscape regarding technology integration in PjBL, identify emerging trends, as well as uncover potential research areas that have not yet been fully explored.

## III. RESULTS AND DISCUSSION

The research results were obtained with the help of the Publish or Perish (PoP) application. Publish or Perish helps researchers find and collect as many as 200 articles related to project based learning models. The articles that researchers collected were articles from 2015 to 2024. There have been several disseminations of research regarding project based learning in the last 10 years. Data on the distribution of project based learning research in the last 10 years can be seen in table 1 below.

**Table 1.** Data on the Distribution of Project Based Learning

| Year | Amount | Percentage (%) |
|---|---|---|
| 2015 | 15 | 7.5 |
| 2016 | 22 | 11 |
| 2017 | 20 | 10 |
| 2018 | 14 | 7 |
| 2019 | 32 | 16 |
| 2020 | 39 | 19.5 |
| 2021 | 37 | 18.5 |
| 2022 | 14 | 7 |
| 2023 | 5 | 2.5 |
| 2024 | 2 | 1 |
| Amount | 200 | 100 |

The data in table 1 was collected using the Publish or Perish application, by entering the keyword "project based learning model". The collected data is saved in CSV and RIS format. Data in CSV format is then managed and analyzed by the number of studies that have examined project based learning. The data is grouped based on the year the article was published. In table 1 you can see several studies regarding project based learning that have been researched by previous researchers. There were 15 studies regarding project based learning in 2015 with a percentage of 7.5%. The most research regarding project based learning was in 2020 with a total of 39 studies and a percentage of 19.5%. You then followed in 2021 with a total of 37 studies and a percentage of 18.5%. The least research data regarding project based learning is in 2024, namely only 2 articles with a percentage of 1%. If depicted in diagram form, the visualization can be seen in Figure 1 below.

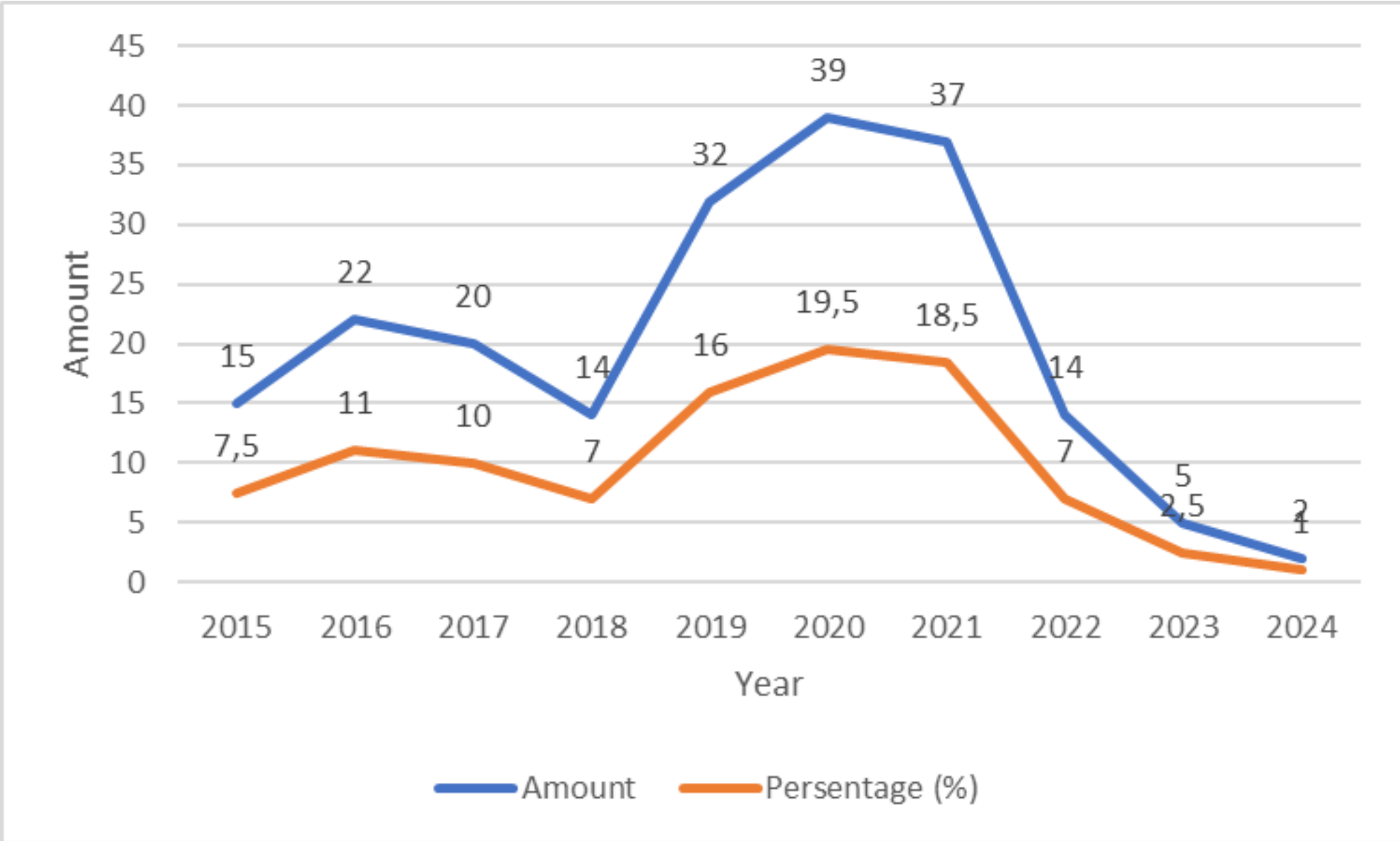

**Fig. 1** Diagram of Distribute Project Based Learning Models

In Figure 2, you can see a graph of the distribution of project based learning models research. 2020 was the year that had the highest distribution of project based learning research with a percentage of 19.5%. In 2015 there were already researchers researching project based learning with a percentage of approximately 7.5%. Research on project based learning models continues to develop today. The percentage of distribution of project based learning models research which was very stable was in 2019, 2020, and 2021 with percentages of 16%, 19.5%, and 18.5% respectively. In 2022, project based learning research will experience a slight decline with a percentage of 7%.

The data stored in RIS format is then entered into the Mendeley application to sort articles according to keywords. A total of 200 article data with the keyword project based learning were analyzed by Mendeley one by one. This activity is used to sort articles that match and do not match keywords. If there are articles that do not match the keywords, the researcher will eliminate the article. The analyzed data is then entered into the VOSViewer application to map the articles. This mapping can be described in three images, namely Networking Visualization, Overlay Visualization, and Density Visualization. In Networking Visualization, data is distributed and depicted in the form of a network. This network is defined as the relationship of one variable with other variables. For Overlay Visualization the data is mapped in several years, and depicted in the form of color brightness levels. The brighter the colors shown in the visualization overlay, the more recently researched the research is. Meanwhile, in Density Visualization, the data is depicted in the form of density and density of the research carried out. There are several variables depicted in Density Visualization using light colors and opaque colors. The brighter the color of the research variable, the more researchers have researched that variable. The results of Networking Visualization on the project based learning model can be seen in Figure 2 below.

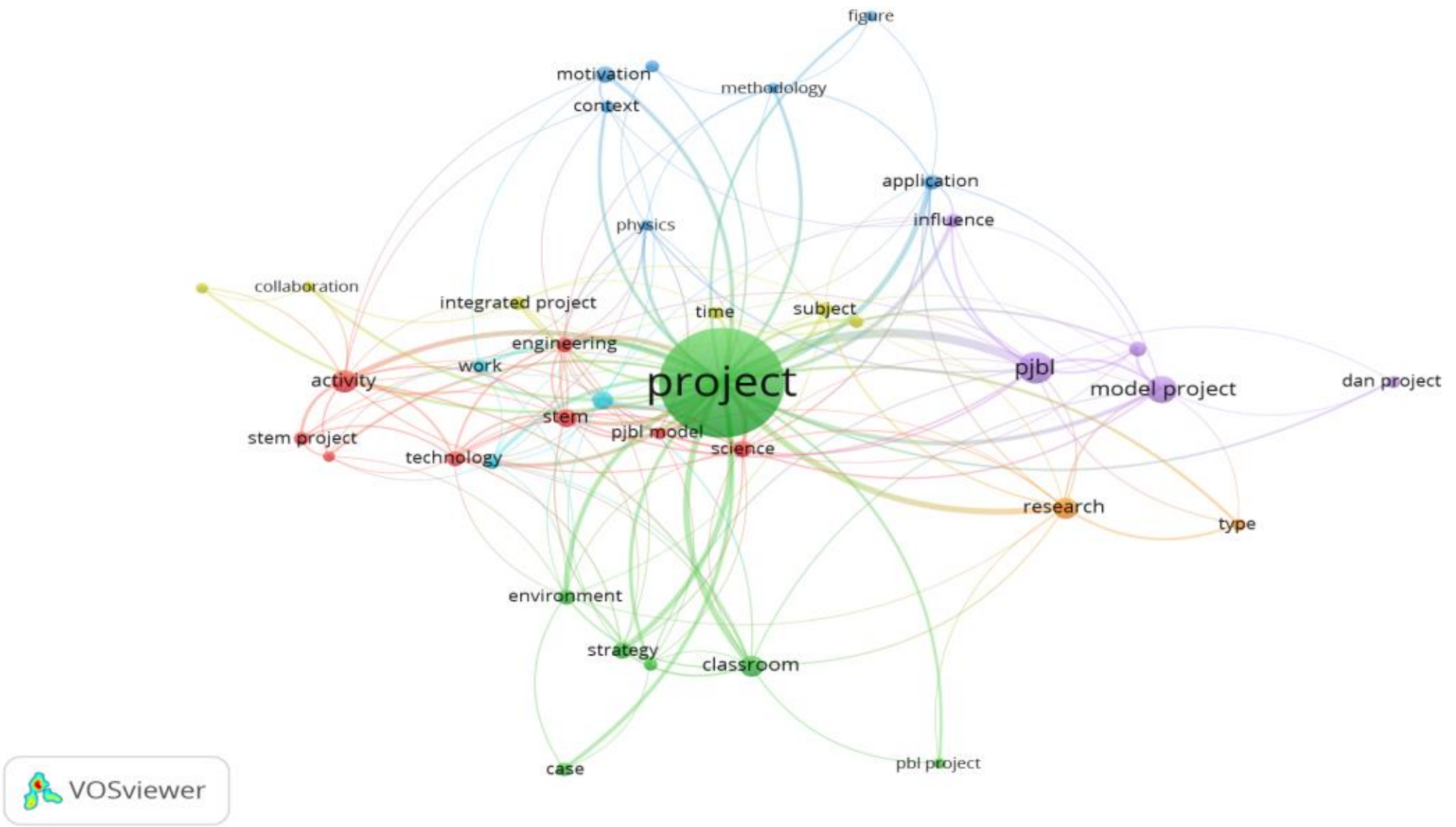

**Fig. 2** Network Visualization of Project Based Learning Model

Network visualization in project based planning can be seen in Figure 1. In the network visualization, you can see the relationships between variables that have been studied by previous researchers. In Figure 1, you can see that every variable that has been studied is depicted in a circle with a different color. Each circle and color has a variety of meanings. In Figure 1 you can see what variables have been studied by researchers. There are several studies that have examined project based learning with technology, with stems, with physics, and with engineering.

Overlay Visualization spreads research variables over the last ten years, namely from 2015 to 2024. Overlay Visualization describes the distribution based on the year the article was published. Each research variable is symbolized by a colored circle and is connected to each other depicted in the form of a straight line. In the overlay visualization it is depicted in the form of dark and light colors. The brighter the color of the variable circle, the more recently the study was researched. A depiction of the overlay visualization can be seen in Figure 3 below.

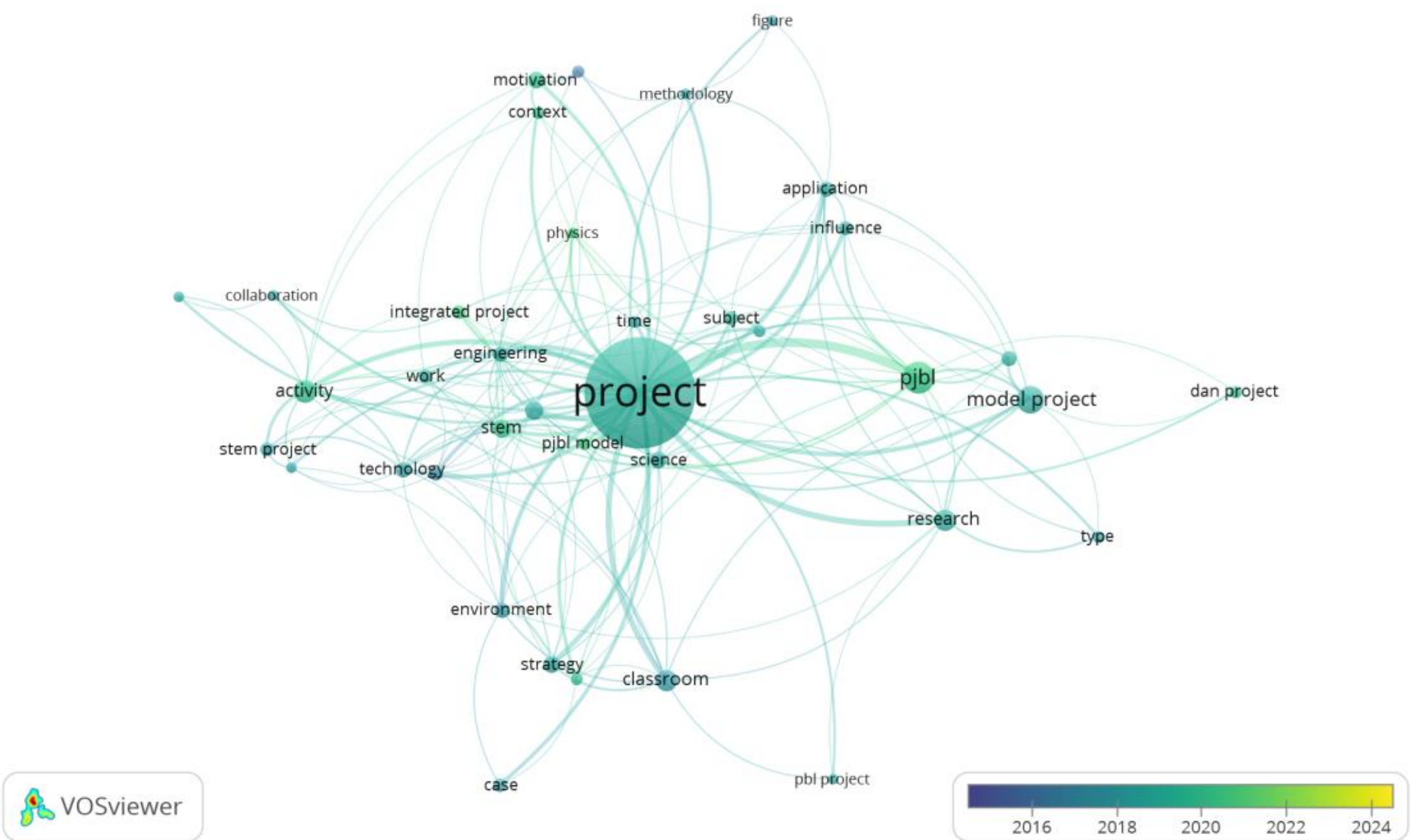

**Fig. 3** Overlay Visualization of Project Based Learning Model

In Figure 2, you can see the distribution of project based learning model research from 2016 to 2024. The brighter the circle color in the visualization overlay, the more recent the research researched. In Figure 2, you can see the latest research regarding project based learning, namely motivation, engineering, and even STEM. Even research on project based learning models with technology is still a trend.

Density Visualization is described in the form of density and research density. Where the more variables are studied, the denser the circle shape of the variables. Density Visualization is described by the light and dark shapes of a variable circle. Density Visualization in the project based learning model can be seen in Figure 4 below.

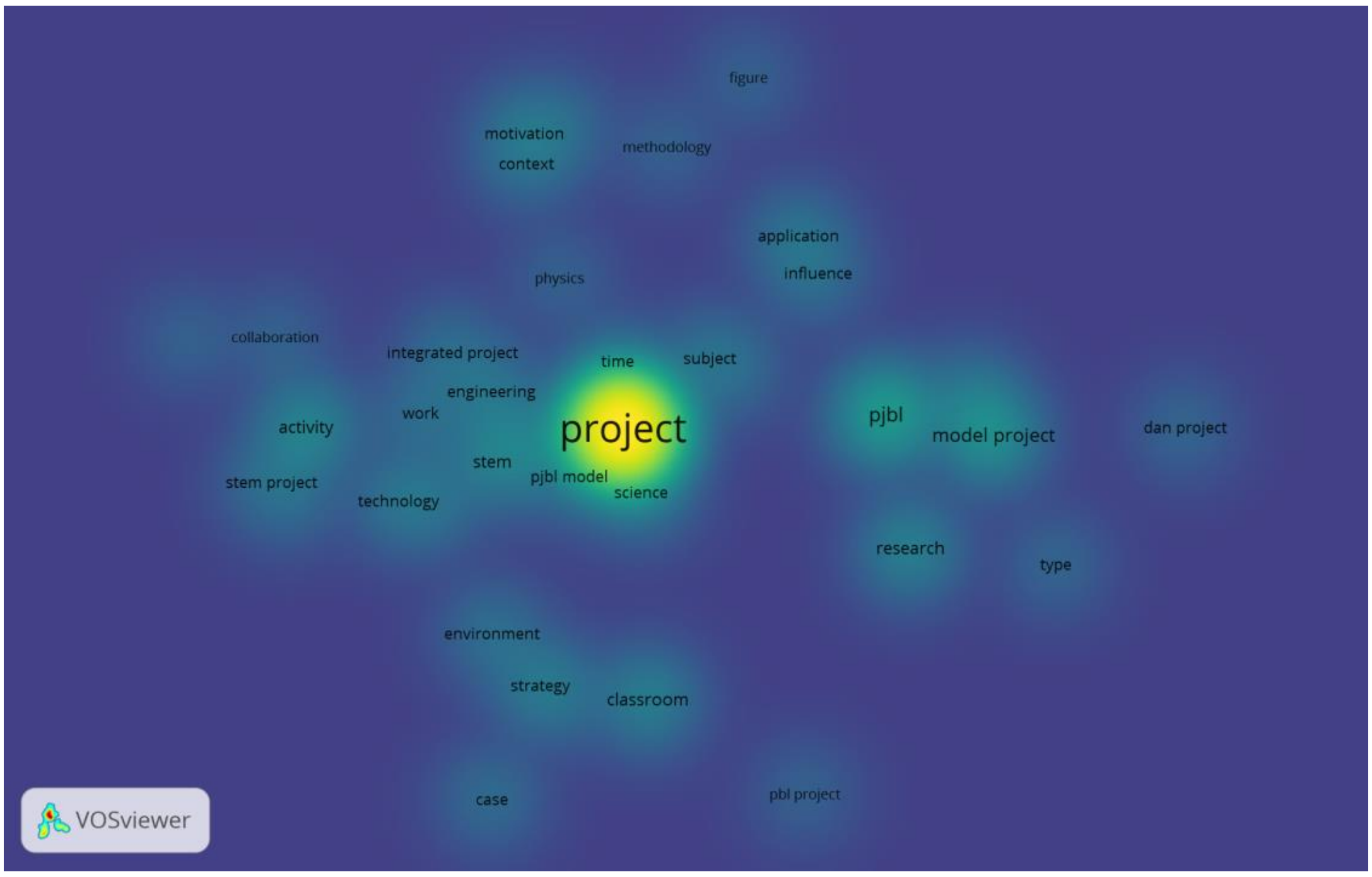

**Fig 4.** Density Visualization of Project Based Learning Model

Figure 4 describes the deployment of the project based learning model. Where the project based learning variables appear to have very bright colors. This means that project variables have been widely studied by several previous researchers. The colors that look quite bright are also about technology, student motivation, even stem and science. It can be stated that research involving the relationship between project based learning models and technology and student motivation is still rarely studied by researchers.

## IV. CONCLUSION

Research regarding the integration of technology into the project based learning model needs to be improved to determine the level of success and motivation of students in mastering learning. Based on research that has been carried out starting from data collection, data processing, carrying out bibliometric mapping using VOS Viewer software, analyzing and describing, the results of bibliometric analysis show that the development of project based model research by integrating technology is still a trend for research.